\documentclass[prd,twocolumn,showpacs,amssymb,superscriptaddress,nofootinbib]{revtex4}
\usepackage{amsmath}
\usepackage{graphicx,epsfig}
\usepackage{dcolumn}
\usepackage{bm}
\usepackage{color}

\newcommand{\ba}{\begin{array}}
\newcommand{\ea}{\end{array}}
\newcommand{\bd}{\begin{displaymath}}
\newcommand{\ed}{\end{displaymath}}
\newcommand{\be}{\begin{equation}}
\newcommand{\ee}{\end{equation}}
\newcommand{\bea}{\begin{eqnarray}}
\newcommand{\eea}{\end{eqnarray}}
\newcommand{\sla}[1]{/\!\!\!#1}

\def\a{\alpha}

\def\b{\beta}

\def\l{\lambda}

\def\q2 {q^2}

\def\r {\rightarrow}

\def\rslep {\tilde{e_R}}

\def\snu {\tilde{\nu}}
\def\lslep {\tilde{e_L}}
\def\stau {\tilde{\tau}}
\def\mer {m_{\rslep}}
\def\mmr {m_{\tilde{\mu}_R}}
\def\mml {m_{\tilde{\mu}_L}}
\def\mel {m_{\lslep}}

\def\bt{\begin{table}}
\def\et{\end{table}}

\begin{document}

\vspace{-0.2cm}
\begin{flushright}
\hspace*{5.5in}
\mbox{RECAPP-HRI-2010-010}
\end{flushright}
\title{Multi-photon signal in supersymmetry comprising non-pointing photon(s)
at the LHC}
\author{Sanjoy Biswas}
\email{sbiswas@hri.res.in}
\affiliation{Harish-Chandra Research Institute,\\
Chhatnag Road, Jhunsi, Allahabad-211019, India.}
\author{Joydeep Chakrabortty}
\email{joydeep@hri.res.in}
\affiliation{Harish-Chandra Research Institute,\\
Chhatnag Road, Jhunsi, Allahabad-211019, India.}
\author{Sourov Roy}
\email{tpsr@iacs.res.in}
\affiliation{Department of Theoretical Physics and Centre for 
Theoretical Sciences, Indian Association for the Cultivation of Science,\\
2A \& 2B Raja S.C. Mullick Road, Kolkata-700032, India.}

\date{\today}

\begin{abstract}
We study a distinct supersymmetric signal of multi-photons in association with jets 
and missing transverse energy. At least one of these photons has the origin in displaced 
vertex, thus delayed and non-pointing. We consider a supersymmetric scenario in which 
the gravitino is the lightest supersymmetric particle (LSP) (with a mass $\sim 1~\mbox{keV}$) 
and the lightest neutralino is the next-to-lightest supersymmetric particle (NLSP). The NLSP 
decays dominantly into a photon and a gravitino within the detector with a decay length 
ranging from $c\tau_{\tilde{\chi}}\sim$ 50-100 cm. In addition, we assume that the second 
lightest neutralino and the lightest neutralino are nearly degenerate and this leads to a 
prompt radiative decay of the next-to-lightest neutralino into a photon and a lightest neutralino 
with a large branching ratio. Such degenerate neutralinos can be realised in various 
representations of the $SU(5)$, $SO(10)$, and $E(6)$ Grand Unified Theories (GUTs). The non-pointing photons can be reconstructed 
at the electromagnetic calorimeter of the ATLAS inner-detector, which have been designed with good
timing and directional resolution. We find that with a centre-of-mass energy $E_{cm}=14 ~\mbox{TeV}$ at 
an integrated luminosity of 100 $fb^{-1}$ one may see evidence of hundreds of tri-photon events and a 
few four-photons events at the LHC, in addition to several thousands di-photon events. We also predict 
the event rates even at the early phase of LHC run. 

\end{abstract}


\maketitle
     
\section{Introduction}
In this era of the Large Hadron Collider (LHC) the TeV scale physics is expected to be probed. 
Supersymmetric Standard Model (SSM) is one of the most interesting and attractive 
candidate for physics beyond the Standard Model (SM). It offers a possibility of gauge coupling 
unification and dark matter candidate, and also solves the gauge hierarchy problem. Once supersymmetry 
(SUSY) is realized as a local symmetry \cite{wess-bagger}, it predicts the existence of the gravitino ${\tilde G}$ 
as the spin-3/2 superpartner of the graviton. Supersymmetry breaking leads to a non-zero mass of the gravitino
through the super-Higgs mechanism, in which the gravitino ``eats up" the spin-1/2 goldstino associated with
spontaneously broken local supersymmetry \cite{super-higgs1,super-higgs2,super-higgs3,super-higgs4}. 
The mass $m_{\tilde G}$ of the gravitino is governed by the scale of SUSY breaking and can range from 
as low as eV scale to as high as 100 TeV scale \cite{nilles,s-p-martin,dine-nelson-shirman,dine-nelson-nir-shirman,
giudice-rattazzi,randall-sundrum,giudice-luty-murayama-rattazzi, buchmuller-hamaguchi-kersten}.
In this work we choose a phenomenological supersymmetric scenario in which gravitino is the lightest 
supersymmetric particle (LSP) with a mass $m_{\tilde G} \sim$ 1 keV and look at the collider signatures of
such a scenario at the LHC. Such light gravitinos also have implications in cosmology. First of all, one should
note that the dark matter relic density is presently known to be $\Omega_{DM}h^2 \simeq$ 0.11 \cite{wmap}. In addition, 
constraints on structure formation require that the bulk of the dark matter should be cold or warm 
\cite{cold-warm-DM}. For a gravitino with a mass $m_{\tilde G} \sim$ 1 keV, nonstandard cosmology and a 
nonstandard gravitino production mechanism are required to satisfy small-scale-structure constraints and to
avoid overclosure \cite{feng-kamionkowski-lee}. One might also need some other dark matter particle. An example of
a nonstandard early-Universe physics is to consider a low-reheating temperature \cite{low-reheat1,low-reheat2}. In 
Ref.\cite{low-reheat2} a low-reheat scenario has been proposed in which a gravitino of mass $m_{\tilde G}$ 
= 1--15 keV can have the right abundance to be the warm dark matter.
  
The interactions of the gravitino are suppressed by the reduced Planck Scale $M_P = 2.4 \times 10^{18}$ GeV and
a light gravitino interacts more strongly than a heavy gravitino. Light gravitinos are primarily produced at 
colliders in the decays of the NLSP. In our scenario the lightest 
neutralino (${\tilde \chi}^0_1$), which is predominantly a bino, is the NLSP and it decays dominantly into a photon 
and a gravitino. These photons are delayed and non-pointing as they are not pointing to the interaction vertex where 
the NLSP is produced. Along with this we also look into the radiative decay of the second-lightest neutralino 
(${\tilde \chi}^0_2$) i.e., $\tilde{\chi}_2^0\rightarrow \tilde{\chi}_1^0 \gamma$, where the emitted photons are prompt. 
Thus our main goal in this paper is to study the spectacular multi-photon events at the LHC where there is a mixture of 
prompt photons and non-pointing photons in the final states. In order to have a large branching ratio of the decay 
$\tilde{\chi}_2^0\rightarrow \tilde{\chi}_1^0 \gamma$, we choose a framework where the $U(1)$ and $SU(2)$ gaugino soft 
SUSY breaking mass parameters $M_1$ and $M_2$, respectively are very close and result in nearly mass degenerate 
${\tilde \chi}^0_2$ and ${\tilde \chi}^0_1$. 

In a minimal supergravity like framework (mSUGRA) the gaugino masses are 
unified at the high scale (unification scale). When they run down to 
electroweak symmetry breaking scale (EWSB) the gaugino mass ratio gets modified through 
renormalisation group effects (RGEs). At the EWSB scale the approximate ratio of the gaugino masses 
are given as $M_1:M_2:M_3\simeq1:2:6$, where $M_3$ is the $SU(3)$ gaugino soft SUSY breaking mass parameter and 
$M_1,M_2$ have been defined in the previous paragraph. So it is very clear from the above ratios that in a mSUGRA 
scenario it is almost impossible to have nearly degenerate neutralions at the EWSB scale. But if the gauginos masses 
are non-universal at the high scale with $M_1>M_2$ then the RGEs compensate for $M_2$ and one can have nearly degenerate 
gauginos at the EWSB scale. In this paper we point out a few grand unified gauge symmetry breaking patterns where this 
feature can be grabbed. 

Light gravitino and its collider signatures have been studied extensively in various context 
\cite{feng-kamionkowski-lee,stump,dimopoulos,thomas,bagger,ambrosanio,ghosal,datta,roy,feng-moroi,hinchliffe,abreu,mercadante,
rimoldi, allanach-lola-sridhar,pagliarone,nojiri,hayward,hamaguchi,smith,wagner,martyn,klasen,shirai,mattia,shirai-yanagida,
chen-adams} and mostly in connection with gauge mediated supersymmetry breaking (GMSB) \cite{dine-nelson-shirman,
dine-nelson-nir-shirman,giudice-rattazzi}. Signatures involving photons are characteristics of scenarios with 
neutralino-NLSP. In most of the cases studied so far the lightest neutralino is predominantly a bino and the second 
lightest neutralino is dominated by its wino component with a large mass splitting between them. However, as emphasized
earlier, we will consider a scenario where the lightest and the second lightest neutralino are approximately degenerate 
in mass. This will lead to multi-photon signatures at the LHC for a 1 keV gravitino, where in the final states we can have 
combinations of prompt and delayed photons. This is a spectacular signal free from Standard Model backgrounds and has 
not been studied earlier. The signature of two non-pointing photons is very much distinct and clean with a large event rate 
at the LHC. We discuss the di-, tri-, and four-photon signals at 14 TeV center-of-mass (CM) energy with 100 fb$^{-1}$ 
integrated luminosity. We find it very hard to get any significant event rates for four photons at 7 TeV CM energy with 
3 fb$^{-1}$, and it is not a surprise. Let us note in passing that triphoton signatures of Randall-Sundrum model have been 
studied recently in Ref.\cite{sudhir-atwood}. 

\begin{figure}[h]
\centerline{\epsfig{file=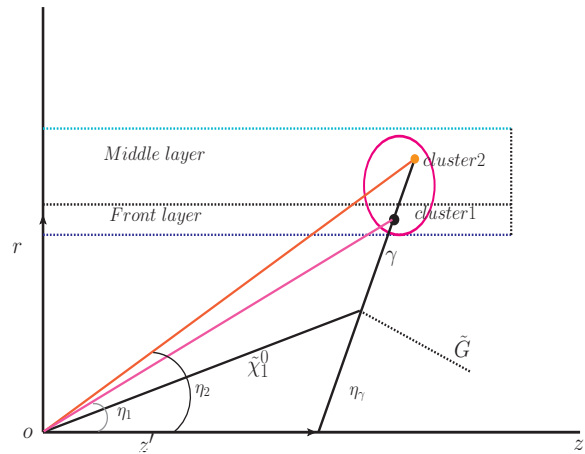, width=8.0cm,height=6.0cm}}
\caption{\label{nlsp}Decay kinematics of the NLSP 
(the lightest neutralino) in the ATLAS detector \cite{hayward,TDR}.}
\label{atlas-tdr}
\end{figure}

We discuss the $p_T$ distributions of multi photons for different suggested 
benchmark points (BP). We use the decay kinematics of the neutralino with a sufficiently long lifetime. Schematic diagram 
of a neutralino decaying into a gravitino and a photon in the ATLAS detector is shown \cite{hayward} in Fig. \ref{atlas-tdr}.
If the decay length of the ${\tilde \chi}^0_1$ is comparable to the size of the ATLAS inner-detector 
\cite{hayward,TDR}, high-$p_T$ photons could enter the calorimeter at angles ($\eta_\gamma$) deviating significantly from 
the nominal angle from the interaction point to the calorimeter cell ($\eta_1$).  

The plan of the paper is as follows. We discuss the gravitino production from $\tilde{\chi}_1^0$ decay in Sec. II.
In Sec. III we discuss how nearly degenerate gaugino masses enhance the branching ratio of the radiative decay of the 
next-to-lightest neutralino and suggest the possible high scale scenarios from where this degeneracy condition can 
be achieved. We suggest four benchmark points for our numerical analysis and these are discussed in Sec. IV. 
In Sec. V we 
start with our goal for collider simulation and discuss the multi-photons+$jet$s associated with missing transverse 
energy (MET) as a potential signal at the LHC. Sec. VI contains the results of our numerical analysis and the conclusion is provided in Sec. VII. 
\section{Gravitino production from neutralino decay }
As discussed earlier, the gravitino gets a mass by the super-Higgs mechanism. The mass of the gravitino is related to the 
fundamental supersymmetry-breaking scale $\sqrt{F}$, as
\bea
m_{\tilde G} = \frac{F}{\sqrt{3}M_P} \simeq 240~{\rm eV} \left[ \frac{\sqrt{F}}{10^3~{\rm TeV}} \right]^2.
\eea
The weak-scale gravitino has a very feeble interaction and thus it is usually hard to find its signatures 
in collider experiments. However, once SUSY is broken spontaneously, the extremely weak gravitino interactions are 
enhanced at energy scales much larger than the gravitino mass $m_{\tilde G}$. This is because in the high energy limit the
gravitino has the same interaction as the goldstino and the couplings of the goldstino are proportional to 
$1/F$ \cite{goldstino1,goldstino2,goldstino3}. Hence the decays of heavier sparticles to gravitinos are faster for light gravitinos. 
The partial decay widths of $\tilde{\chi}_1^0$ to $\tilde{G}$ 
are given as \cite{bagger,allanach-lola-sridhar,ibarra-buchmuller}:
\begin{eqnarray}
\Gamma(\tilde{\chi}_1^0\rightarrow \gamma \tilde{G})&=&\frac{k_{1\gamma}}{48\pi}
\frac{m_{\tilde{\chi}_1^0}^5}{M_P^2 m_{\tilde{G}}^2};\\
\Gamma(\tilde{\chi}_1^0\rightarrow Z \tilde{G})&=&\frac{2k_{1Z_{T}}+k_{1Z_{L}}}{96\pi}
\frac{m_{\tilde{\chi}_1^0}^5}{M_P^2 m_{\tilde{G}}^2} [1-\frac{m_Z^2}{m_{\tilde{\chi}_1^0}^2}]^4;\\
\Gamma(\tilde{\chi}_1^0\rightarrow \phi \tilde{G})&=&\frac{k_{1\phi}}{96\pi}
\frac{m_{\tilde{\chi}_1^0}^5}{M_P^2 m_{\tilde{G}}^2} [1-\frac{m_{\phi}^2}{m_{\tilde{\chi}_1^0}^2}]^4,
\end{eqnarray}
where,
\begin{eqnarray}
k_{1\gamma} &=&|N_{11}\cos \theta_w + N_{12} \sin \theta_{w}|^2, \nonumber \\
k_{1Z_{T}} &=&|N_{11}\sin \theta_w + N_{12} \cos \theta_{w}|^2, \nonumber \\
k_{1Z_{L}}&=&|N_{13}\cos \theta_{\beta} - N_{14} \sin \theta_{\beta}|^2, \nonumber \\
k_{1h^0}&=&|N_{13}\sin \alpha - N_{14} \cos \alpha|^2, \nonumber \\
k_{1H^0}&=&|N_{13}\cos \alpha + N_{14} \sin \alpha|^2, \nonumber \\
k_{1A^0}&=&|N_{13}\sin \beta + N_{14} \cos \beta|^2.
\end{eqnarray}
Here $N_{ij}$ are the neutralino mixing matrices, $\theta_{w}$ is the weak mixing angle, 
$\alpha$ is the Higgs ($\Phi=h^0,H^0,A^0$) mixing angle and $\tan\beta$ is the ratio of the
vacuum expectation values of the two Higgs doublets $H_1$ and $H_2$ in the SSM. 
From the above expressions it is clear that for a bino-like NLSP $N_{11}\cos \theta_{w}$ 
is much larger than $N_{12} \sin \theta_{w}$. The decay modes into the 
photon dominates over $Z$ and $\phi$ channels as the later two have phase-space suppressions.
Assuming that the decay widths in $Z$ and $\phi$ channels are negligible, the decay length
of the lightest neutralino is given by 
\bea
c\tau_{\tilde \chi} = \frac{1}{k_{1\gamma}}\left(\frac{100~{\rm GeV}}{m_{{\tilde \chi}^0_1}} \right)^5 
\left( \frac{\sqrt{F}}{100~{\rm TeV}} \right)^4 \times 10^{-2}~{\rm cm}.
\eea
For a pure bino-like lightest neutralino and $m_{\tilde G}$ = 1 keV, we get a decay length 
$c\tau_{\tilde \chi} \approx$ 70 cm.
\section{Radiative decay of neutralino: $\tilde{\chi}_2^0 \rightarrow \tilde{\chi}_1^0 \gamma$}
The radiative decay of second lightest neutralino emanates at the one-loop level and 
decay width is given as \cite{hw}:
\bea
\Gamma(\tilde{\chi}_2^0 \to \tilde{\chi}_1^0\gamma) = 
 \frac{g^2_{\tilde{\chi}_2^0 \tilde{\chi}_1^0\gamma}}{8\pi} 
 \frac{(m_{\tilde{\chi}_2^0}^2-m_{\tilde{\chi}_1^0}^2)^3}{m_{\tilde{\chi}_2^0}^5},
\label{rad_decay-width} 
\eea
where $g_{\tilde{\chi}_2^0 \tilde{\chi}_1^0\gamma} \propto eg^2/16\pi^2$ is an effective coupling. 
This radiative decay is enhanced \cite{hw, am} by a kinematic factor when $\tilde{\chi}_1^0$ and 
$\tilde{\chi}_2^0$ are nearly degenerate as in this regime of parameter space three body decays are suppressed 
by a factor $\epsilon^5$, where $\epsilon=(1-m_{\tilde{\chi}_1^0}/m_{\tilde{\chi}_2^0})$. 
It is being noted in \cite{am, Diaz, bk} that the decay branching ratio of 
$\tilde{\chi}_2^0 \rightarrow \tilde{\chi}_1^0 \gamma$ is much larger for 
large $\tan \beta$ and $\mu > M_1,M_2$ with $|\mu| \sim M_1 \tan \beta/2$. In the general MSSM scenario 
the radiative decay branching ratio can reach nearly 100$\%$ \cite{bk}  for $|M_1|,\ M_2 \lesssim 1000$ GeV, 
with $|M_1|,\ M_2<|\mu |$ and $|M_1|\sim M_2$. We have calculated the branching ratio 
of the radiative decay of the second lightest neutralino using {\bf \begin{footnotesize}SDECAY version 1.3b \end{footnotesize}} \cite{sdecay}. 

\subsection{Radiative decay with non-universal gaugino masses}
In \cite{bk}, the enhancement conditions in the radiative decay branching 
fractions are justified for the minimal supergravity(mSUGRA) models with 
non-universal gaugino masses. The part of the N=1 supergravity Lagrangian
(the part that contains only the real part of the left-chiral superfields $\Phi_{i}$)
containing the kinetic energy and the mass terms for the gauginos and the gauge bosons 
can be written as:
\bea
 e^{-1} {\mathcal{L}}&=& \nonumber \\
&-&\frac{1}{4} Re f_{\alpha \beta}(\phi)(-1/2
\bar{\lambda}^
{\alpha} D\!\!\!\!/ \lambda^{\beta})-\frac{1}{4} Re f_{\a \b}(\phi)F_{\mu
  \nu}^{\a}F^{\beta \mu \nu}  \nonumber \\
 & + &\frac{1}{4} e^{-G/2}
G^{i}((G^{-1})^{j}_{i})[\partial f^*_{\a \b}(\phi^*)/\partial
{\phi^{*j}}]\l^{\a}\l^{\b} + h.c \nonumber \\
\label {lag}
\eea
where $G^i = \partial{G}/\partial{\phi_{i}}$ and $(G^{-1})^{i}_j$ is the
inverse matrix of ${G^j}_i\equiv \partial{G} / {\partial{\phi^{*i}}\partial{\phi_j}}$,
$\l^{\a}$ is the gaugino field, and $\phi$ is the scalar component of the 
chiral superfield $\Phi$ and $F_{\mu \nu}^{\a}$ is the unified gauge kinetic term. 
The $F$-component of the symmetry breaking scalar field $\Phi$ generates 
gaugino masses with a consistent SUSY breaking with non-zero vacuum expectation value 
(vev) of the chosen $\tilde{F}$, where 
\bea
\tilde{F}^{j}=\frac{1}{2} e^{-G/2} [G^{i}((G^{-1})^{j}_{i})].
\eea
The $\Phi^{j}$'s can be a set of GUT singlet supermultiplets 
$\Phi^{S}$, which are part of the hidden 
sector, or a set of non-singlet ones $\Phi^N$, fields associated 
with the spontaneous breakdown of the GUT group to 
$SU(3)\otimes SU(2)\otimes U(1)$. The non-trivial gauge kinetic 
function $f_{\a\b}(\Phi^{j}) $ can be expanded in 
terms of the non-singlet components of the chiral superfields in the 
following way 
\bea
f_{\alpha \beta}(\Phi^{j})= f_{0}(\Phi^{S})\delta_{\a\b}+\sum_{N}\xi_{N}(\Phi^s)
\frac{{\Phi^{N}}_{\a\b}}{M_P}+ {\mathcal{O}}(\frac{\Phi^N}{M_P})^2, 
\nonumber \\
\eea
where $f_0$ and $\xi^N$ are functions of chiral singlet superfields,
essentially determining the strength of the interaction and
$M_P$ is the reduced Planck mass$=M_{Pl}/\sqrt{8\pi}$.\\
The contribution to the gauge kinetic function
from  $\Phi^{N}$ has to come through symmetric products of the 
adjoint representation of the associated GUT group, since $f_{\a\b}$ 
has such transformation property for the sake of gauge 
invariance. The non-universal gaugino masses are calculated for $SU(5)$, $SO(10)$ 
and $E(6)$ grand unified gauge groups in \cite{nonug}. 
The results for the ratios of gaugino masses are given in Table \ref{table:E662R}. 
We have tabulated here only the cases where $M_1$ and $M_2$ are nearly degenerate 
at the EWSB scale with $M_1<M_2$. This fits in our scenario.
\begin{center}
\begin{table}
\begin{tabular}[c]{|c|c|c|}
\hline
\phantom{X}Representations\phantom{X} &  \phantom{x} $M_1$ : $M_2$ : $M_3$ \phantom{x} & \phantom{x} 
$M_1$ : $M_2$ : $M_3$ \phantom{x}\\
 & \phantom{x} (at $M_{\rm GUT}$) \phantom{x} & \phantom{x} (at $M_Z$) \phantom{x} \\
\hline
{\bf 75 $\subset SU(5)$}  & $-5$ : $3$ : $1$ & $-5$ : $6$ : $6$
\\
{\bf 210, 770 $\subset SO(10)$} &  &
\\
\hline
{\bf 2430 $\subset E(6)$}   & $-\frac{9}{5}$ : $1$ : $1$ & $-1.8$ : $2$ : $6$
\phantom{$\frac{\hat I}{\hat I_1}$} 
\\
 & $\frac{5}{2}$ : $-\frac{3}{2}$ : $1$ & $2.5$ : $-3$ : $6$
\\
\hline
\end{tabular}
\caption{\small \it{Ratios of gaugino masses for $F$-terms in representations of
$SU(5)$, $SO(10)$ and $E(6)$ leading to nearly degenerate 
gauginos at low scale.\label{table:E662R}}}
\end{table}
\end{center}

\section{Benchmark points}
In this section we present four benchmark points (see Table \ref{tab:2}) we have worked with 
to demonstrate that the nearly degenerate $M_1$ and $M_2$ at the EWSB scale can lead to radiative 
decay of the second lightest neutralino. In addition to this we have also shown that if one has 
$M_1<M_2$ and the gravitino in the bottom of the spectrum, then the lightest neutralino 
has a sizeable branching fraction to decay into a photon and gravitino. This leads to
multi-photon signatures in collider experiment. The spectrum has been generated using
the {\bf \begin{footnotesize}SuSpect version 2.41 \end{footnotesize}}\cite{SUSPECT} with
all the input parameters specified at the electroweak scale. The gravitino mass is taken
to be $\sim 1$ keV which is necessary for the fact that the lightest neutralino decays
within the detector with a decay length $c\tau_{{\tilde \chi}^0_1} \sim$ 50--100 cm. The radiative 
decay of the $\tilde{\chi}_2^0$ and decay of $\tilde{\chi}_1^0$ have been calculated 
using {\bf \begin{footnotesize}SDECAY version 1.3b \end{footnotesize}}\cite{sdecay}. The benchmark 
points we have worked with are consistent with all the low energy constraints like muon 
$(g-2)_{\mu}$, $b\r s\gamma$ and the LEP limit on the lightest Higgs boson mass and other charged 
particles masses \cite{Amsler:2008zzb,constraints}. 

Throughout all of our benchmark points we have kept the value of $m_{\tilde{\chi}^0_2}$ and $m_{\tilde{\chi}^0_1}$ 
nearly the same with different choices of $\mu$, $\tan\beta$, squarks, gluino and slepton masses.
The high value of $\mu$ is important for enhancement of the radiative decay branching fraction
of $\tilde{\chi}_2^0$ into a ${\tilde \chi}^0_1 \gamma$ pair. We have worked with $m_{\tilde{g}}$ starting from
as low as 413 GeV to 740 GeV. We set $A_t=A_\tau=A_b=A_0$ = -1000 GeV. The large value of $|A_0|$ is required for 
obtaining a large radiative decay branching ratio (BR) of $\tilde{\chi}_2^0$. For $A_0$= 0, the three-body-decay 
modes of $\tilde{\chi}_2^0$ are dominant and the radiative decay is very much suppressed in our case. We have also noted 
that the radiative decay branching fraction depends less significantly on the sign of $A_0$. It is a little less for the 
positive value of $A_0$ than the negative one keeping $|A_0|$ same. In Table \ref{tab:3} we tabulate the decay branching 
fraction of the $\tilde{\chi}_2^0$ and $\tilde{\chi}_1^0$ for our choice of input parameters. From this table one can see 
the effect of $\mu$, $\tan\beta$, squark and slepton masses on the radiative decay of $\tilde{\chi}_2^0$. However, the
decay branching fraction of the lightest neutralino is determined once the mass of the gravitino, the lightest
neutralino and the neutralino mixing parameters are fixed and does not depend at all on the choices of 
squarks, gluino and sleptons masses.
\begin{table}[htbp]
\footnotesize
\begin{tabular}{||c||c|c|c|c||}
\hline
\hline
      & {\bf BP-1}&{\bf BP-2}&{\bf BP-3}&{\bf BP-4} \\
\hline
$\tan \beta$ & 40  & 15 & 10 & 15 \\
$\mu$   & 1500 & 1500  & 1500  & 2500   \\
\hline
$\mel,\mml$  &601 & 601 & 502 & 701 \\
$\mer,\mmr$  &601 & 601 & 502& 701 \\
$m_{\snu_{e_L}},m_{\snu_{\mu_L}}$&597 &596 & 496& 697 \\
$m_{\snu_{\tau_L}}$&597 & 596& 496 & 697\\
$m_{\stau_1}$&591 & 567& 473 & 652\\
$m_{\stau_2}$&611 &634 & 529 & 747 \\
\hline
$m_{\tilde{\chi}^0_1}$&200 & 199& 206& 206\\
$m_{\tilde{\chi}^0_2}$&236 &237 & 236 & 239\\
$m_{\tilde{\chi}^{\pm}_1}$&236 &237  & 236& 240 \\
$m_{\tilde{g}}$&413  & 414 & 688 & 739\\
$m_{\tilde{d}_L}$&613 & 614& 521 & 728 \\
$m_{\tilde{d}_R}$&611 & 612& 518& 727 \\
$m_{\tilde{u}_L}$&609 & 609 &515 & 724 \\
$m_{\tilde{u}_R}$&610 & 610 & 516 & 725 \\
$m_{\tilde{b}_1}$&599 & 573 & 486 & 680 \\
$m_{\tilde{b}_2}$&626  & 651  & 551  & 771 \\
$m_{\tilde{t}_1}$&366 & 421 & 215 & 422 \\
$m_{\tilde{t}_2}$&735  & 708 & 627 & 434 \\
$m_{h^0}$ &110  & 118 & 115 & 119  \\
\hline
\hline
\end{tabular}\\
\caption {\small \it Proposed benchmark points (BP) for the 
study of radiative decay of $\tilde{\chi}^0_2$ and the NLSP $\tilde{\chi}^0_1$. 
We have set $A_0=-1000$ \mbox{GeV} for the third generation squarks
and sleptons and it is zero for the rest. 
Masses of the particles and $\mu$ are given in \mbox{GeV}.}
\label{tab:2}       
\end{table}

\begin{table}[htbp]
\begin{tabular}{||c||c|c|c|c||}
\hline
\hline
 & {\bf BP-1}  & {\bf BP-2}  & {\bf BP-3}  & {\bf BP-4} \\
\hline
$\tilde{\chi}_2^0\rightarrow \tilde{\chi}_1^0 \gamma$  & 0.30  & 0.11   &  0.26   &  0.10   \\ 
\hline
$\tilde{\chi}_1^0 \rightarrow \gamma \tilde{G}$  & 0.89  & 0.89  &  0.87  &  0.88    \\
\hline 
\hline
\end{tabular}
\caption{\small \it {Branching fractions for the decays 
$\tilde{\chi}_2^0\rightarrow \tilde{\chi}_1^0 \gamma$ and 
$\tilde{\chi}_1^0 \rightarrow \gamma \tilde{G}$ for different benchmark points.}}

\label{tab:3}
\end{table}

\section{Collider simulation}
\begin{figure}
\includegraphics{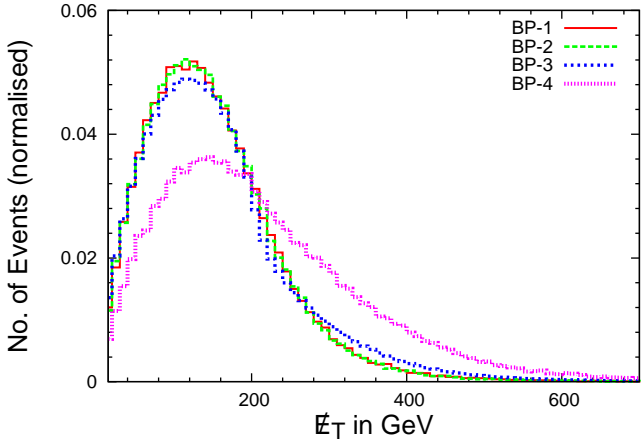}
\caption{Missing Energy distribution for different benchmark points
with $E_{cm}$={\rm 14} {\rm TeV}.}
\label{missing-et}
\end{figure}

The $\tilde{\chi}^0_1$ and $\tilde{\chi}^0_2$ are produced in cascade decays of squarks and gluinos 
accompanied by hard jets. In an $R$-parity conserving scenario the gravitino is produced
at the end of each cascade, which goes undetected at the collider detector, leading to large amount 
of missing transverse energy ($\sla E_T$) (see, Fig. \ref{missing-et}). Thus one can have multi-photon signals in 
association with hard jets and $\sla E_T$. The collider simulation has been done with a centre of mass energy $E_{cm}$=14 
TeV, at an integrated luminosity of 100 fb$^{-1}$ using the event generator {\bf \begin{footnotesize} 
PYTHIA 6.4.16\end{footnotesize}} \cite{PYTHIA}. A simulation for the early LHC run at $E_{cm}$=7 TeV and 
integrated luminosity of 3 fb$^{-1}$ has also been performed. We have used the parton distribution function 
CTEQ5L \cite{Lai:1999wy} with the factorisation ($\mu_F$) and renormalisation ($\mu_R$) scale set at 
$\mu_R =\mu_F =$average mass of the final state particles produced in the initial hard scattering. The effects of 
Initial and Final State Radiation (ISR/FSR) have also been taken into account. Below we mention the numerical values 
of various parameters used in our calculation \cite{Amsler:2008zzb}\\
\noindent
$M_Z=91.187$ GeV, $M_W=80.398$ GeV, $M_t=172.3$ GeV, 
$\a^{-1}_{em}(M_Z)=127.9$, $\a_{s}(M_Z)=0.118$, 
where $M_Z$, $M_W$ and $M_t$ are the masses of the $Z$-boson, $W$-boson and top quark, 
respectively. $\a_{em}(M_Z)$ and $\a_{s}(M_Z)$ are the electromagnetic coupling 
constant and strong coupling constant respectively at the scale of $M_Z$.

\subsection{Event selection criteria}
\begin{figure}
\includegraphics{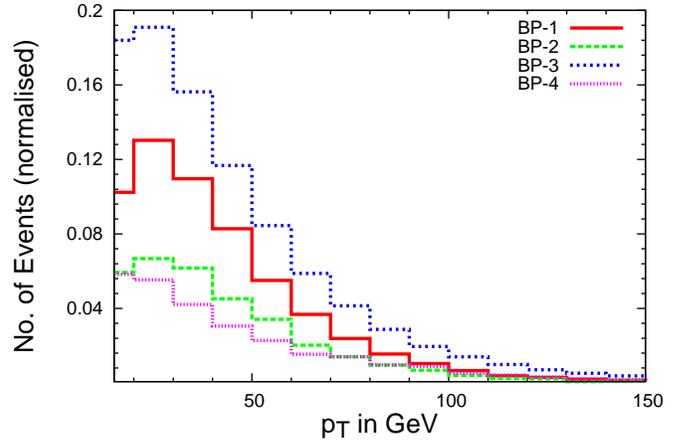}
\includegraphics{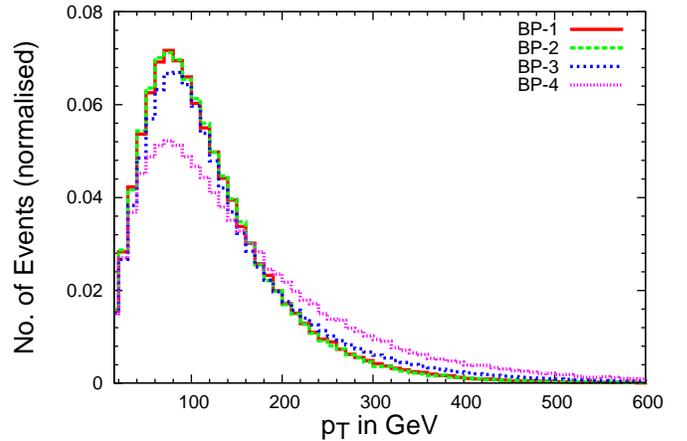}
\caption{$p_T$ distributions of the prompt photon (top) and
 non-pointing photon (bottom) with $E_{cm}$={\rm 14} {\rm TeV} for all benchmark
points.}
\label{photon-pt}
\end{figure}

We have considered the following final states to demonstrate the event rates in
multi-photon channels:
 
\begin{itemize}
\item $2\gamma+\sla E_T+jets$ 
 \item $3\gamma+\sla E_T+jets$
\item $4\gamma+\sla E_T+jets$ 
\end{itemize}

\noindent where at least one of these photons has the origin in displaced vertex due to the fact that
the decay length of the lightest neutralino is $\mathcal{O}$ (50-100 cm). The photon out of
a $\tilde{\chi}^0_2$ decay is soft (see Fig. \ref{photon-pt}-top) while the $p_T$ of the photon coming from a 
$\tilde{\chi}^0_1$ are normally hard (see Fig. \ref{photon-pt}-bottom) as the mass difference between $\tilde{\chi}^0_2$ 
and $\tilde{\chi}^0_1$ is $\mathcal{O}$(30 GeV). 

\begin{figure}
\includegraphics[scale=0.9]{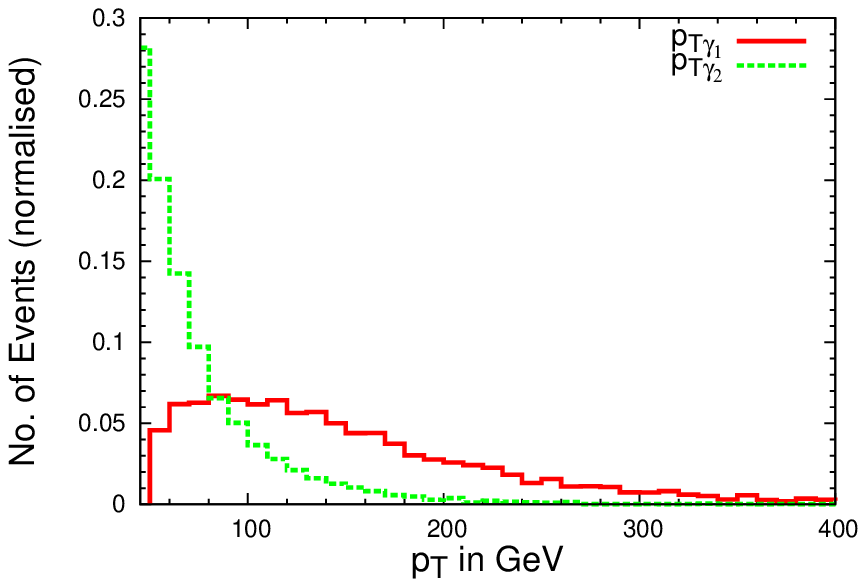}
\includegraphics[scale=0.9]{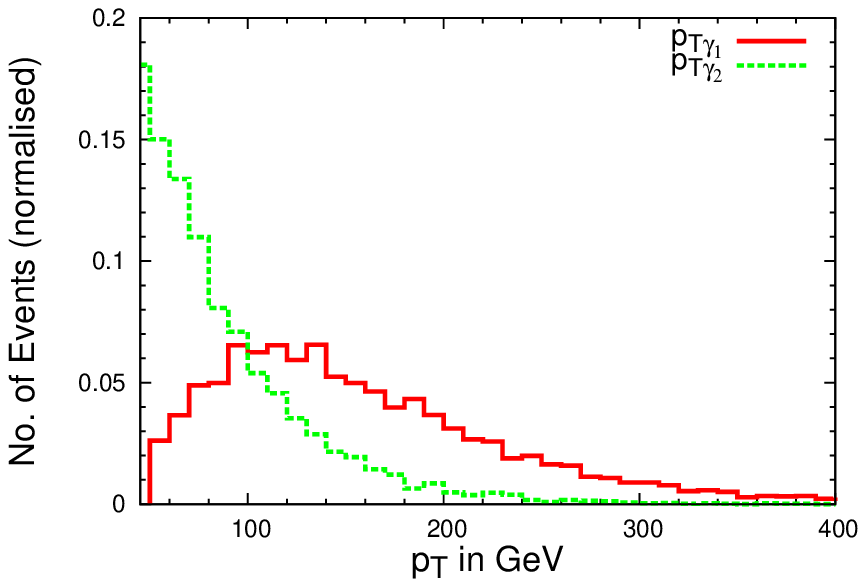}
\includegraphics[scale=0.9]{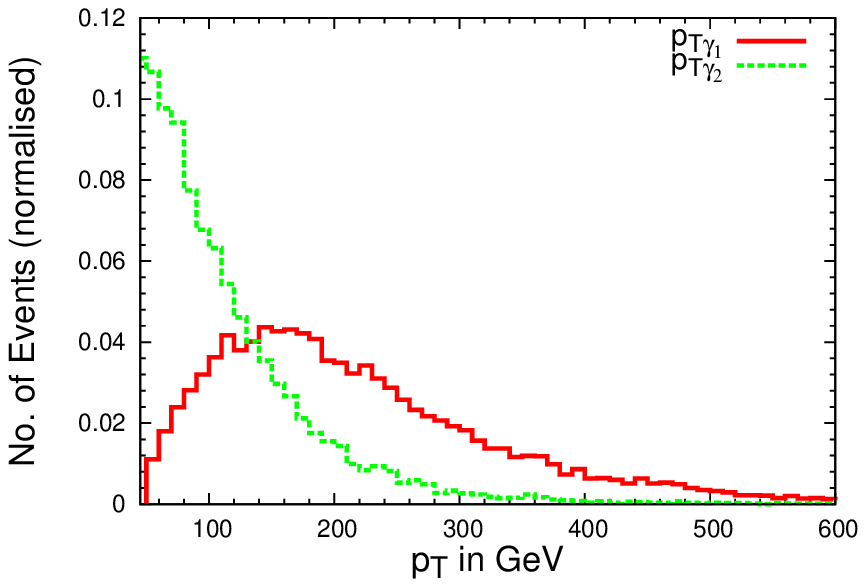}
\includegraphics[scale=0.9]{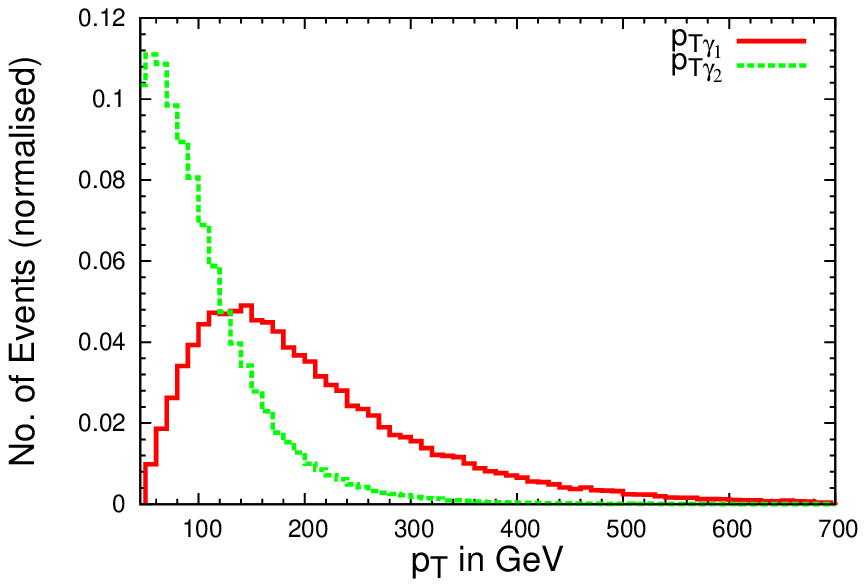}
\caption{$p_T$ distributions of di-photons for (from top to bottom) {\rm BP-1}, {\rm BP-2},
{\rm BP-3}, and {\rm BP-4} with $E_{cm}$={\rm 14} {\rm TeV}.}
\label{photon-pt2}
\end{figure}
\begin{figure}
\includegraphics[scale=0.9]{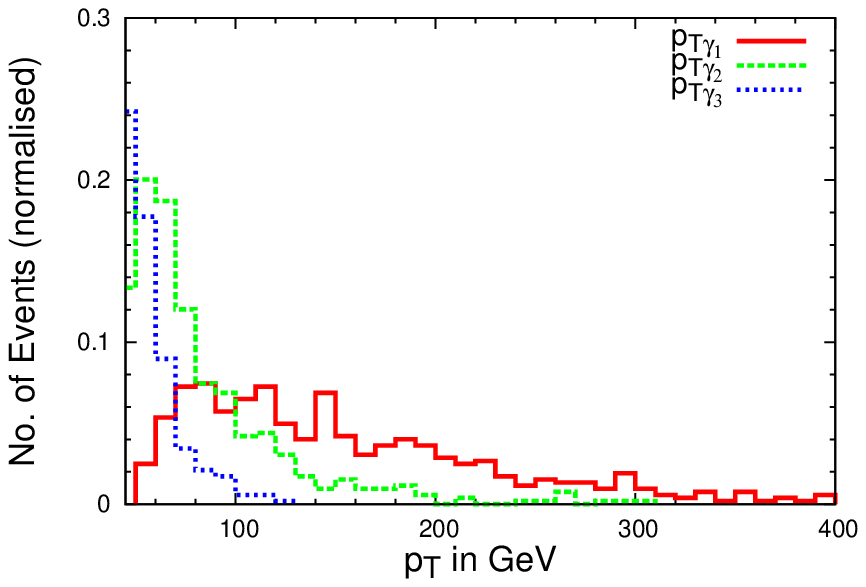}
\includegraphics[scale=0.9]{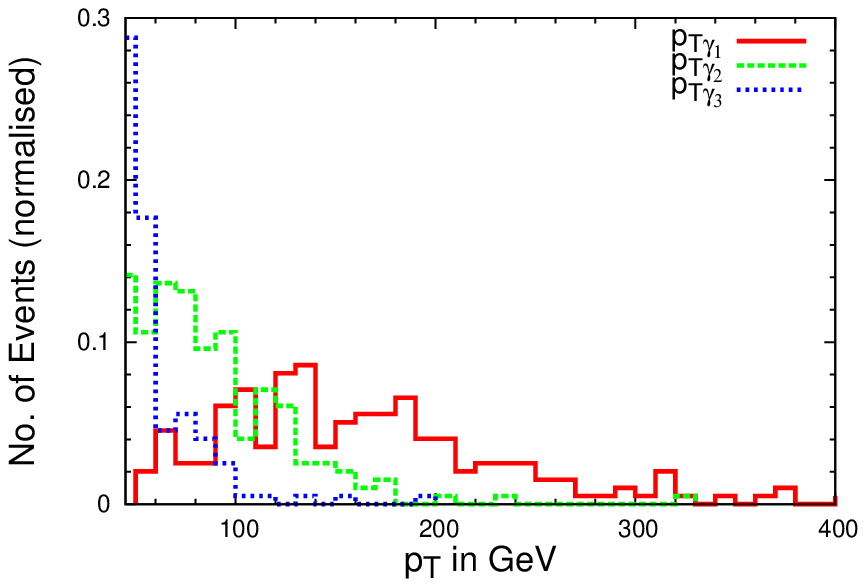}
\includegraphics[scale=0.9]{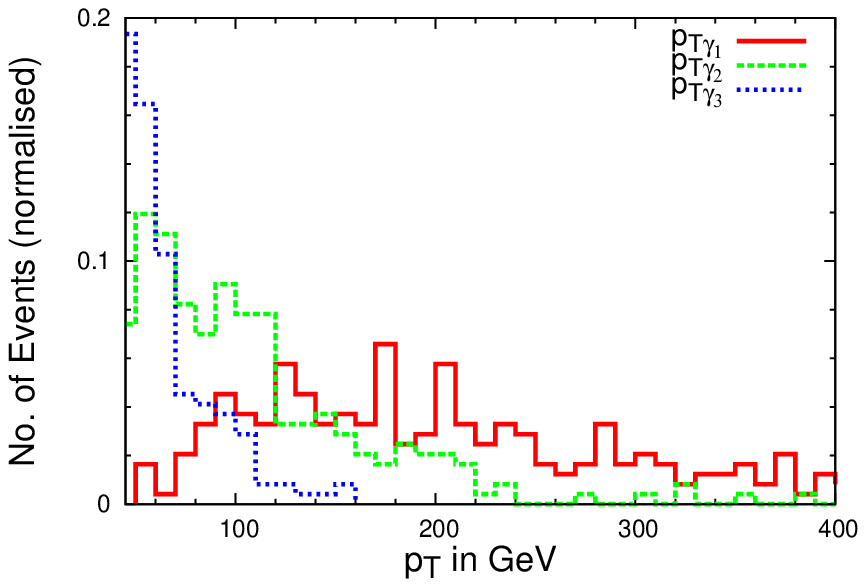}
\includegraphics[scale=0.9]{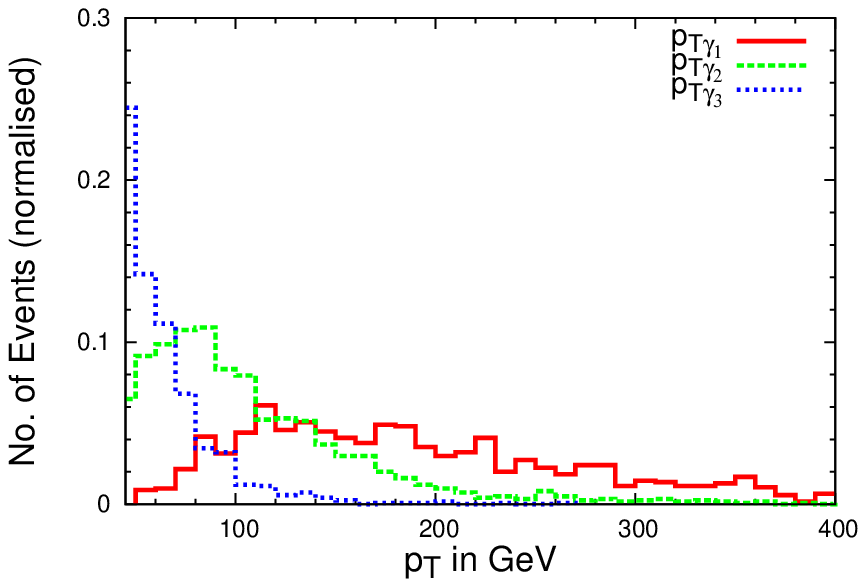}
\caption{$p_T$ distributions of tri-photons for (from top to bottom) {\rm BP-1}, {\rm BP-2},
{\rm BP-3}, and {\rm BP-4} with $E_{cm}$={\rm 14} {\rm TeV}.}
\label{photon-pt3}
\end{figure}
\begin{figure}
\includegraphics[scale=0.9]{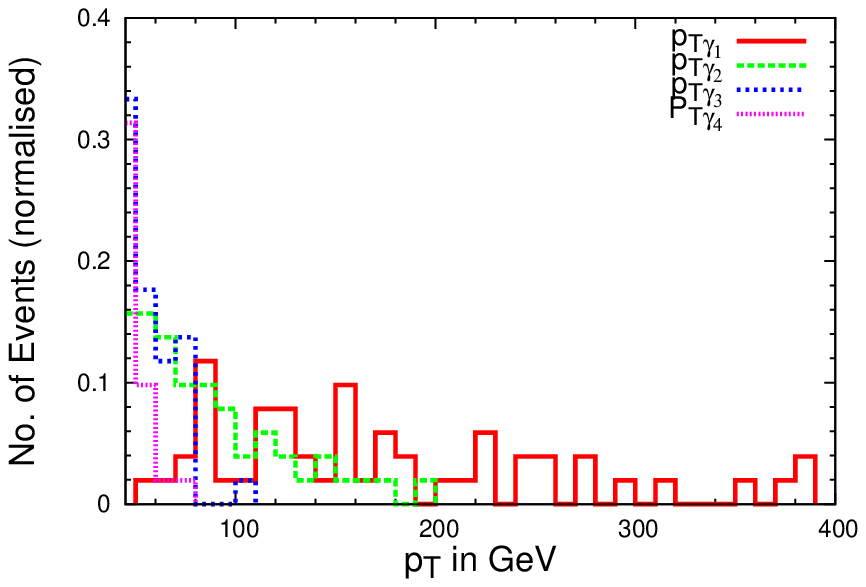}
\includegraphics[scale=0.9]{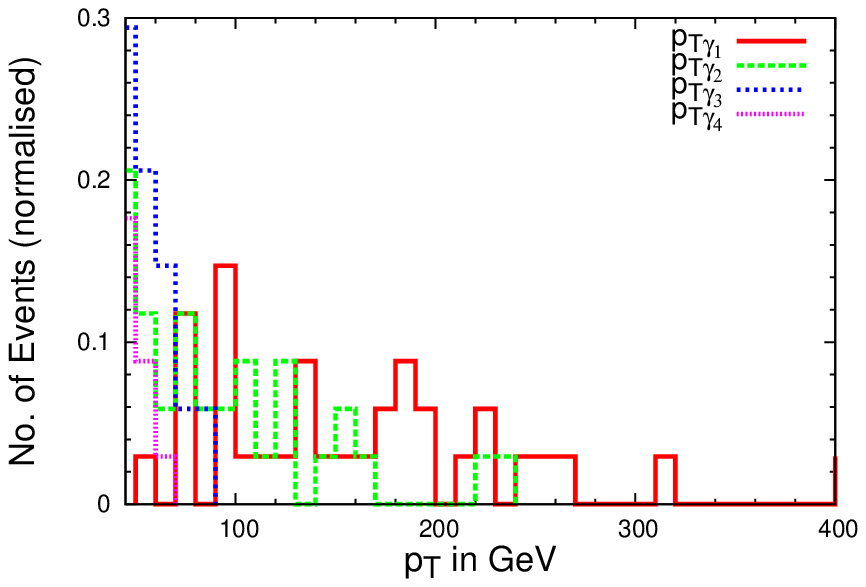}
\includegraphics[scale=0.9]{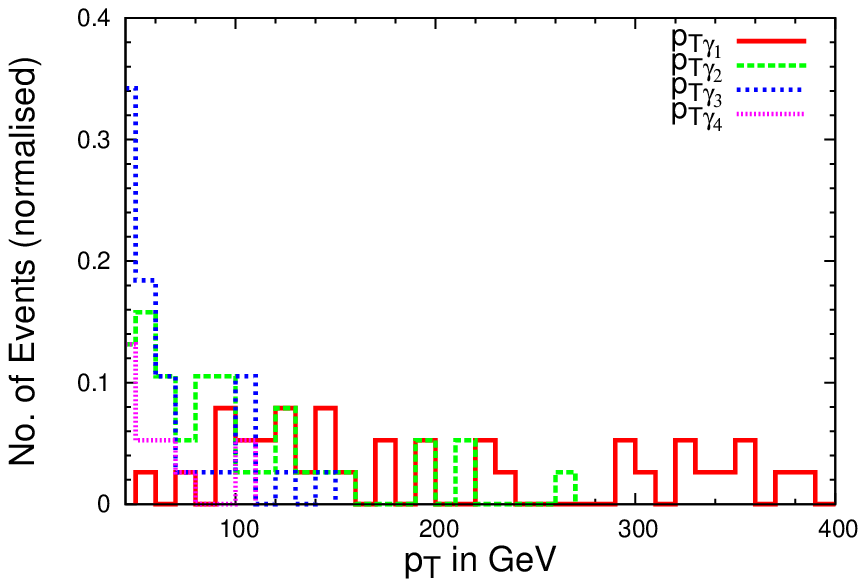}
\includegraphics[scale=0.9]{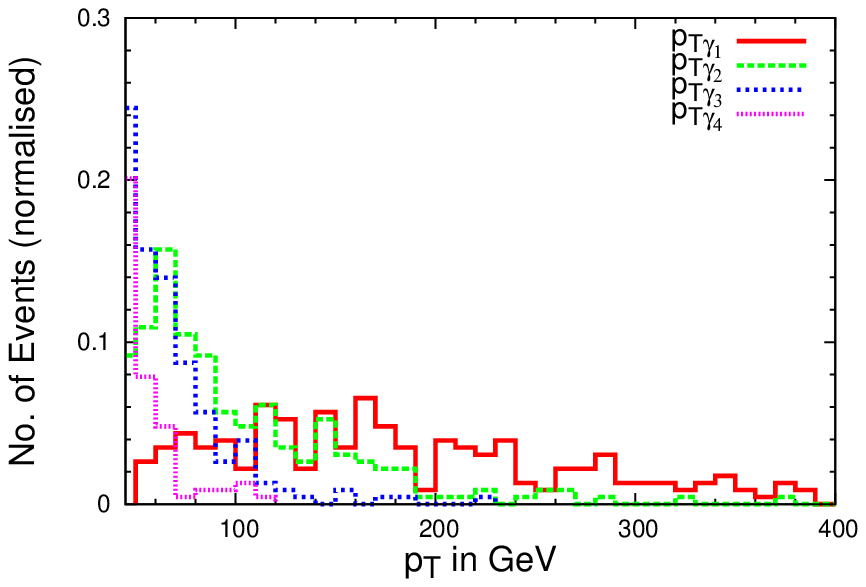}
\caption{$p_T$ distributions of four photons for (from top to bottom) {\rm BP-1}, {\rm BP-2},
{\rm BP-3}, and {\rm BP-4} with $E_{cm}$={\rm 14} {\rm TeV}.}
\label{photon-pt4}
\end{figure}

The following requirements have been implemented to select isolated photons:
\begin{itemize}
 
\item We have identified photons with $p_T$ more than 30 GeV and $|\eta|\leq$2.5.

\item  A minimum $\Delta R$ separation between two photons has been demanded in terms of $\Delta R>0.2$,
where $\Delta R=\sqrt{(\Delta \eta)^2 + (\Delta \phi)^2}$.

\item A lepton-photon and jet-photon isolation of $\Delta R_{l\gamma}>0.4$ and $\Delta R_{j\gamma}>0.6$,
respectively have been imposed.

\item The sum of hadronic $E_T$ deposit in a cone of $\Delta R=0.2$ around the photon is required to be
$\Sigma |E_T|<10 ~\mbox{GeV}$.

\item To reduce the di-photon background from $\pi^0\rightarrow 2\gamma$ we have also required a 
photon-photon invariant mass cut $m_{\pi}-20 ~\mbox{GeV}<M_{\gamma\gamma}<m_{\pi}+20 ~\mbox{GeV}$.
\end{itemize}

The photons have been ordered according to their hardness (see Figs. 4, 5 and 6) and a minimum $p_T$
cut has been imposed on each of them depending on the various final states:

\begin{itemize}
\item {\bf di-photon}: $p_{T_{\gamma_1}}> 50 ~\mbox{GeV}$, $p_{T_{\gamma_2}}> 40 ~\mbox{GeV}$ 

\item {\bf tri-photon}: $p_{T_{\gamma_1}}> 50 ~\mbox{GeV}$, $p_{T_{\gamma_2}}> 40 ~\mbox{GeV}$ , 
$p_{T_{\gamma_3}}> 30 ~\mbox{GeV}$

\item {\bf four-photon}: $p_{T_{\gamma_1}}> 50 ~\mbox{GeV}$, $p_{T_{\gamma_2}}> 40 ~\mbox{GeV}$ , 
$p_{T_{\gamma_3}}> 30 ~\mbox{GeV}$, $p_{T_{\gamma_4}}> 30 ~\mbox{GeV}$.
\end{itemize}



We have also incorporated the probability of jet-faking as photon, which is taken to be 0.1\% 
\cite{terwort,TDR} and an identification efficiency of 60\% has been used for the non-pointing photons
following \cite{TDR,photon-conversion}. We have not taken into account the rapidity dependence of the 
identification efficiency and used a uniform efficiency for a conservative approach.

\section{Results}
In this section we present the numerical results of our simulation. In Table \ref{tab:4} we have
presented the number of events in the multi-photon channels after applying the basic cuts listed
in the previous section. The different benchmark points we have selected correspond to similar
$m_{\tilde{\chi}^0_1}$ and $m_{\tilde{\chi}^0_2}$ with different values of $\mu$, $\tan\beta$, 
slepton, and squark masses as given in Table \ref{tab:2}. The radiative decay branching fraction of 
$\tilde{\chi}^0_2\r \tilde{\chi}^0_1\gamma$ (see Table \ref{tab:3}) depends on the choice of squarks and 
slepton masses as well as on the values of $\tan\beta$ and $\mu$, which in turn affect the event rates in 
various multi-photon channels. In BP-1 the gluino is lighter than the squarks. In this case, $\tilde{\chi}^0_2$ is
produced via radiative and three-body decay of gluino which together has a branching fraction of more
than 50\%. The left-handed squarks decay into a $\tilde{\chi}^0_2 q$-pair either directly (with a branching fraction 
$\sim$ 8\%) or via gluino decay. The right-handed squarks mainly decay into a gluino and a quark pair and
the gluino further can decay into a $\tilde{\chi}^0_2 g$ or $\tilde{\chi}^0_2 q\bar{q}$-pair. The situation is similar
in BP-2 with only difference is that it has smaller radiative decay branching fraction (11\%) of decaying
into a $\tilde{\chi}^0_2\r \tilde{\chi}^0_1\gamma$-pair due to small $\tan\beta=10$. In BP-3 the squarks are lighter 
than the gluino. In this case the gluino directly decays into a $q {\tilde q}$-pair. Therefore the production cross-section
of $\tilde{\chi}^0_2$ in SUSY cascade decreases as the dominant contribution in this case comes only from the decay
of left-handed squarks with a branching fraction ranging from 30\%-35\%. The radiative decay branching 
fraction $\tilde{\chi}^0_2\r \tilde{\chi}^0_1\gamma$ is slightly greater than BP-2, due to the fact that the 
squarks and slepton masses are smaller than that in BP-2 (see Table \ref{tab:2}), which contribute in the loop.

Above all, due to different squarks and gluino masses at different benchmark points the overall SUSY production
cross-section changes from one benchmark point to the other. This combined with the different decay branching 
fractions of ${\tilde q}_L \r {\tilde \chi}_2^0 q$ for various benchmark points, affects the production cross-section 
of the second lightest neutralino in cascade decay of squarks and gluino and shows up in the final event rates.

\begin{table}[htbp]
\begin{tabular}{||c||c|c|c|c||}
\hline
\hline
{\bf SIGNAL} & {\bf BP-1}  & {\bf BP-2}  & {\bf BP-3}  & {\bf BP-4} \\
\hline
$2\gamma+\sla E_T+jets$  & 7942  & 6604  & 8150  & 9549     \\ 
\hline
$3\gamma+\sla E_T+jets$  & 597 & 220  & 165  & 162   \\ 
\hline
$4\gamma+\sla E_T+jets$  & 8  & 2  & 3  &  3  \\
\hline 
\hline
\end{tabular}
\caption{\small \it {Number of signal events, after applying the basic cuts at 
an integrated luminosity of \mbox{100 fb}$^{-1}$ and the center of mass energy of \mbox{14 TeV} 
for all our benchmark points.}}

\label{tab:4}
\end{table}

From Table \ref{tab:4} one can find that in the di-photon channel one has substantial rate at $E_{cm}=14$ 
TeV and integrated luminosity of 100 fb$^{-1}$. The dominant contribution to this channel comes from the 
two non-pointing photons out of a $\tilde{\chi}^0_1\r \tilde{G}\gamma$ decay from the two cascades, and constitutes
of more than 92\% of the total di-photon cross-section. The decay branching fractions of $\tilde{\chi}^0_1\r \tilde{G}
\gamma$ are more than 87\% for all of our benchmark points. The next sub-dominant contribution to it comes from
one non-pointing photon from $\tilde{\chi}_1^0$ decay and the other prompt photon from radiative decay of $\tilde{\chi}_2^0$.
This constitutes $\sim 7\%$ of the total di-photon cross-section. The rest comprises of two prompt photons
when we have radiative decay of $\tilde{\chi}_2^0$ from both the cascade or a combination of prompt or non-pointing 
photon with the ISR/FSR photon, the fraction of which is rather small. We have also presented the di-photon
rates even at the early phase of LHC run with $E_{cm}=7$ TeV at an integrated luminosity of 3 fb$^{-1}$ (see
Table \ref{tab:5}). In BP-1, the di-photon rate is larger than BP-2 with nearly identical spectrum. This attributes 
to the fact that the radiative decay branching fraction of ${\tilde \chi}^0_2$ at BP-2 is one-third of that in BP-1.

\begin{table}[htbp]
\begin{tabular}{||c||c|c|c|c||}
\hline
\hline
{\bf SIGNAL} & {\bf BP-1}  & {\bf BP-2}  & {\bf BP-3}  & {\bf BP-4} \\
\hline
$2\gamma+\sla E_T+jets$  & 20 &17  & 19  &  17   \\ 
\hline
$3\gamma+\sla E_T+jets$  & 2  &  1  & 0  &  0   \\ 
\hline 
\hline
\end{tabular}
\caption{\small \it {Number of signal events in the di-photon channel, after applying the basic 
cuts at an integrated luminosity of \mbox{3 fb}$^{-1}$ and the center of mass energy of \mbox{7 TeV} 
for all our benchmark points.}}

\label{tab:5}
\end{table}


The number of events in the tri-photon channel are relatively small since one of these photon comes
from radiative decay of $\tilde{\chi}_2^0$. The overall tri-photon event rates are small due to following two reasons:
the smaller radiative decay branching fraction of the second lightest neutralino and together with the fact that 
the photon out of a $\tilde{\chi}_2^0$ decay comes with relatively small $p_T$ (see Fig. \ref{photon-pt3}), because 
of small mass splitting between $m_{\tilde{\chi}^0_1}$ and $m_{\tilde{\chi}^0_2}$. Hence in very small fraction of events 
they pass the requisite hardness cut. The effect is much more sever in case of four-photon channels as one can see from 
Table \ref{tab:4}. Though we have quoted the event rates also in this case since one hardly has any contamination from the
SM backgrounds, one still needs higher luminosity for better statistics.

\section{Summary and conclusion}
We have considered a supersymmetric scenario in which the gravitino (with a mass $\sim$ 1 keV) is the LSP and
the NLSP is the lightest neutralino. The second lightest neutralino is nearly degenerate in mass with the 
lightest neutralino. A possible origin of such a degeneracy at the low-scale lies in the form of non-universal 
high scale ($\sim 10^{16}$ GeV) inputs of the soft SUSY breaking gaugino mass parameters. We have pointed out that 
such non-universal high-scale inputs can be realised in various representations of the $SU(5)$, $SO(10)$, and $E(6)$ 
GUT group. 

We have examined the decays of the NLSP and the second lightest neutralino at the LHC. In such a scenario the second 
lightest neutralino has a substantial branching fraction of decaying into a photon and the lightest neutralino. The 
branching fraction depends on $\mu$, $\tan\beta$, and other scalar masses in the theory. The lightest neutralino is 
predominantly a bino and it too decays into a photon and a gravitino with a large branching ratio. Thus one naturally 
has spectacular multi-photon final states in a collider experiment, where light neutralinos are produced in abundance. 
The photons out of the NLSP decay are non-pointing and can be identified 
in the ATLAS inner-detector with an efficiency of 60\%. Such non-pointing photons are free from any SM contamination.

We have studied the di-photon, tri-photon, and four-photon final states in association with hard jets and 
missing transverse energy in the context of LHC both at $E_{cm}=7 ~\mbox{TeV}$ and $14~\mbox{TeV}$ and at an integrated 
luminosity of 3 fb$^{-1}$ and 100 fb$^{-1}$, respectively. Though the di-photon and tri-photon signals look promising, 
one needs higher luminosity for the four-photon case. 

Detection of such multi-photon final states comprising non-pointing photons at the LHC would have serious implications
for early-Universe cosmology and supersymmetry model building. On one hand one needs to have a suitable supersymmetry
breaking mediation mechanism, which allows for light gravitino with a mass $\sim$ 1 keV and nearly mass degenerate 
${\tilde \chi}^0_2$ and ${\tilde \chi}^0_1$. On the other hand, this may give some hints towards a non-standard 
cosmological scenario leading to a keV gravitino which is a warm dark matter candidate with right relic abundance.

\vspace{0.2 cm}
\noindent
{\large {\bf Acknowledgment}}
 
\vspace{0.2 cm}
SB and JC would like to thank the Department of Theoretical Physics, Indian Association for the Cultivation 
of Science for the hospitality where a part of the work was done. SB also thanks KEK and the Institute of Physics and 
Mathematics of the Universe for their hospitality while part of this work was being carried out. This work was partially 
supported by funding available from the Department of Atomic Energy, Government of India for the Regional Centre for 
Accelerator-based Particle Physics, Harish-Chandra Research Institute, XIth Plan `Neutrino physics'. Computational work 
for this study was partially carried out at the cluster computing facility of Harish-Chandra Research Institute 
({\tt http:/$\!$/cluster.mri.ernet.in}).

\end{document}